\documentclass[%
 reprint,
nofootinbib,
 amsmath,amssymb,
 aps,
]{revtex4-2}

\usepackage{graphicx}
\usepackage{dcolumn}
\usepackage{bm}
\usepackage[usenames,dvipsnames]{xcolor}
\usepackage[normalem]{ulem} 
\usepackage[hidelinks]{hyperref}

\newcommand{\mbf}[1]{\mathbf{#1}}

\DeclareMathAlphabet\mathbfcal{OMS}{cmsy}{b}{n}
\DeclareMathOperator{\dprime}{{\prime\prime}}

\begin{document}

\preprint{APS/123-QED}

\title{Mitigating antenna gain errors with HyFoReS in CHIME simulations}

\author{Haochen~Wang$^{1,2}$}
\email[]{hcwang96@mit.edu}
\author{Panupong~Phoompuang$^{1, 3}$}
\author{Kiyoshi W.~Masui$^{1,2}$}
\author{Arnab~Chakraborty$^{4,5}$}
\author{Simon~Foreman$^{6}$}
\affiliation{$^1$ Department of Physics, Massachusetts Institute of Technology, 77 Massachusetts Avenue Cambridge, MA 02139, USA}
\affiliation{$^2$ MIT Kavli Institute for Astrophysics and Space Research, Massachusetts Institute of Technology, 77 Massachusetts Avenue Cambridge, MA 02139, USA}
\affiliation{$^3$ Department of Physics, Brown University, 182 Hope Street Providence, RI 02912, USA}
\affiliation{$^4$ Department of Physics, McGill University, 3600 rue University, Montr\'eal, QC H3A 2T8, Canada}
\affiliation{$^5$ Trottier Space Institute, McGill University, 3550 rue University, Montr\'eal, QC H3A 2A7, Canada}
\affiliation{$^6$ Department of Physics, Arizona State University, Tempe, AZ, USA}

\date{\today}

\begin{abstract}
Hybrid Foreground Residual Subtraction (HyFoReS) is a new family of algorithms designed to remove systematics-induced foreground contamination for 21-cm intensity mapping data. Previously, the algorithm was shown to be effective in mitigating beam perturbations in sky maps from the Canadian Hydrogen Intensity Mapping Experiment (CHIME). In this study, we apply HyFoReS to CHIME simulations and test the algorithm's ability to mitigate antenna gain-type systematics in polarized visibilities. Simulating a two-cylinder telescope similar to the CHIME pathfinder, we find that HyFoReS reduces foreground bias caused by bandpass perturbations to a level below the thermal noise, provided that the RMS value of the perturbations is on the order of $10^{-4}$ or lower. When tested with complex antenna-dependent gain errors, HyFoReS can reduce residual foreground bias in the power spectrum by up to three orders of magnitude. While noise bias and second-order perturbations are currently the limiting factors for the algorithm, we have demonstrated that HyFoReS can suppress gain-induced foreground leakage in polarized data from 21-cm telescopes, aiding in the detection of the 21-cm auto-power spectrum for hydrogen intensity mapping experiments.
\end{abstract}

\maketitle

\section{introduction}
 Detecting the 21-cm line emission of neutral hydrogen is a novel way to probe the large-scale structure (LSS) of the Universe. This can be done using the technique of intensity mapping, which observes the accumulated 21-cm signal from many galaxies without resolving each individual source \citep{chang_pen, Loeb}. Compared with traditional galaxy spectroscopic surveys, hydrogen intensity mapping has the advantages of high mapping speed and relatively low cost, allowing experiments to survey large volumes of the Universe through a wide range of cosmic history. 
    
The promise of the field is reflected by the large number of 21-cm experiments that are either underway or being planned. Low redshift ($z < 5$) experiments, including CHIME \citep{chime_overview}, uGMRT \citep{ugmrt}, MeerKAT \citep{meetkat_IM}, CHORD \citep{chord}, HIRAX \citep{hirax}, and BINGO \citep{bingo}, aim to detect the Baryon Acoustic Oscillation (BAO) to constrain the time evolution of dark energy. Experiments at high redshifts ($6< z < 30$), such as HERA \citep{HERA_2017a}, LOFAR \citep{LOFAR_2017}, LWA \citep{LWA}, and MWA \citep{MWA_2019}, and SKA-LOW \citep{SKA_low}, observe the Epoch of Reionization to investigate the effects of the first stars on the Universe's environment. These experiments have produced results in cross-correlations with other LSS surveys \citep{chime_lyman, chime_stacking, GBT_cross, meerkat_cross, Li_2021, Anderson_2018, kiyo_survey, Chang_2010}, upper limits on auto-correlation \citep{Switzer_2013, HERA_limit, ugmrt}, and a claimed detection of auto-correlation \citep{paul2023detection}. 
    
To detect the 21-cm signal, all hydrogen intensity mapping experiments face the challenge of astrophysical foreground contamination, mostly due to the Galactic synchrotron emission and radio point sources, which overwhelm the 21-cm signal by 3 to 5 orders of magnitude in brightness. In principle, foregrounds can be separated from the 21-cm signal with their distinct spectral behaviors---foreground synchrotron emission has a smooth frequency spectrum, but the 21-cm signal intensity varies rapidly across frequencies. This is the basic assumption behind traditional linear foreground filtering methods, such as Karhunen-Loève (KL) \citep{COBE_KL} and delay-based filters \citep{delay_filter}, which were shown to be effective in removing foregrounds in simulations with perfect instruments \citep{first_mmode, mmode}. However, in real experiments, calibration errors and uncharacterized telescope responses imprint complicated frequency dependence on the intrinsically smooth foreground spectrum, thus breaking the assumption of traditional linear foreground filters and causing bright foreground residuals in the filtered data \citep{chime_stacking}. Residual foreground contamination, induced by such systematics, is the biggest challenge for hydrogen intensity mapping experiments.
    
Various new methods have been proposed to mitigate systematics and foreground contamination \citep{Byrne_2021, Sievers_2017, calamity, haochen_first_paper}. In particular, Hybrid Foreground Residual Subtraction (HyFoReS) was introduced in \citep{haochen_first_paper} as a new class of foreground removal algorithms that specifically target systematics-induced foreground residuals. HyFoReS is based on traditional linear filters but adds a cross-correlation step to further remove foregrounds. A linear foreground filter is applied first to the data, giving a rough signal and a foreground estimate. The signal estimate is dominated by residual foregrounds due to systematics. HyFoReS then cross-correlates the signal and foreground estimates to isolate and remove residual foregrounds in the signal estimate.

 
HyFoReS was applied to sky maps from the CHIME telescope in an earlier study \citep{wang2024} to mitigate beam-induced foreground residuals, and improved the signal-to-noise ratio of the CHIME-eBOSS cross-correlation \citep{chime_stacking} by $20\%$. Although this specific case showcased the algorithm's ability at mitigating beam errors in the map space, HyFoReS was designed with a broader goal in mind, that is to mitigate any parametrizable systematics in either map or visibility space. In this study, we introduce gain-type errors to CHIME simulations and apply HyFoReS to remove foregrounds in polarized visibilities. We thus demonstrate that HyFoReS can be a versatile foreground removing method capable of mitigating various types of systematics for 21-cm experiment data.
    
The structure of this paper is as follows. In Sec.~\ref{sec:form}, we first review the formalism of HyFoReS and specify how to tailor it to target gain-type perturbations in CHIME simulations. In Sec.~\ref{sec:sim}, we give a brief overview of the CHIME simulation pipeline and explain how we introduce gain-type perturbations, including bandpass and complex antenna gain errors, to the simulated data. We then apply HyFoReS to remove residual foreground contamination and show the results in Sec.~\ref{sec:result}. In Sec.~\ref{sec:dis}, we compare the cases between bandpass and complex gain perturbations and discuss the limitations of HyFoReS and CHIME simulations. We also highlight the novelties of this study and ideas for future improvement. Finally, we present the conclusions in Sec.~\ref{sec:con}.

\section{formalism} \label{sec:form}

In this study, we use HyFoReS to mitigate antenna gain-type systematics in simulated 21-cm data. To gradually increase the complexity, we start by simulating and mitigating simple bandpass gain errors, which only depend on frequency, before we simulate complex antenna gain perturbations. In Sec.~\ref{subsec:form_rev}, we provide a brief review of HyFoReS (similar to Sec. III A of \citep{wang2024}) and demonstrate how to apply the algorithm to reduce bandpass errors. In Sec.~\ref{subsec:app}, we extend the formalism to address complex antenna perturbations in CHIME simulations.

\subsection{Review of HyFoReS} \label{subsec:form_rev}

HyFoReS was first derived in \citep{haochen_first_paper}. The algorithm relies on a traditional linear foreground filter to obtain a rough signal and foreground estimate. Due to telescope systematics, the signal estimate is dominated by foreground residuals. The algorithm then cross-correlates the signal estimate with the foreground estimate to draw out foreground contamination, and then subtracts it from the signal estimate.

We use $\bm{v_d}$, $\bm{v_{HI}}$, and $\bm{v_F}$ to represent the data, signal, and foregrounds, respectively, in vector form. Note that although $\bm{v_d}$ represents data ``visibilities," the formalism does not change in the map space (as demonstrated in \citep{wang2024}). Assuming telescope systematics are parametrizable and act like multiplicative perturbation factors on the signal and foregrounds, we can model our data as 
\begin{equation} \label{eq:data_form}
    \bm{v_d} = (\mbf{I} + \mbf{G})(\bm{v_{HI}} + \bm{v_F}),
\end{equation}
where
\begin{equation} \label{eq:G_form}
    \mbf{G} = \sum_i g_i \mbf{\Gamma_i}
\end{equation}
is a perturbation matrix that multiplies the signal and foregrounds with a set of systematics $g_i$, and $\mbf{I}$ is the identity matrix. Later, we will use HyFoReS to estimate the systematics $g_i$ and remove their effects from the signal. The matrix $\mbf{\Gamma_i}$ serves as a derivative matrix for each perturbation $g_i$ and determines which subspace of the data each perturbation $g_i$ affects. \footnote{Note that HyFoReS is not designed to mitigate additive systematics such as radio frequency interference (RFI) and excessive noise. We assume that RFI and excessive noise have been identified (for example, by comparing the data with the expected noise level) and masked in the data prior to HyFoReS.}

For example, the observed visibilities in CHIME simulations (described in Sec.~\ref{sec:sim}) are functions of frequency $\nu$, baseline $b$, and right ascension (RA) $\alpha$, so we write the visibilities in index notation as $(v_d)_{\nu, b, \alpha}$. (Note that RA is used as an observing-time coordinate in this paper.) The visibilities in CHIME simulations also depend on polarization, which is suppressed here for simplicity but will become important in Sec.~\ref{subsec:app}. For bandpass perturbations with only frequency dependence, we write $g_i \rightarrow g_\nu$ and $\mbf{\Gamma_i} \rightarrow \mbf{\Gamma_\nu}$ with
\begin{equation} \label{eq:gamma_nu_form}
    (\Gamma_\nu)_{(\nu^\prime b^\prime \alpha^\prime) (\nu^{\prime\prime} b^{\prime\prime} \alpha^{\prime\prime})} = \delta_{b^\prime b^{\prime\prime}} \delta_{\alpha^\prime \alpha^{\prime\prime}} \delta_{\nu^\prime \nu^{\prime\prime}} \delta_{\nu^\prime \nu},
\end{equation}
which means that the bandpass error corresponding to frequency $\nu$ is multiplied into visibilities across all baselines and RA's of that frequency. With this index notation and $\mbf{\Gamma_\nu}$ defined in Eq.~(\ref{eq:gamma_nu_form}), Eq.~(\ref{eq:data_form}) simply becomes
\begin{equation} \label{eq:data_form_index}
(v_d)_{\nu,b,\alpha} = (1 + g_\nu) \left[ (v_{HI})_{\nu,b,\alpha} + (v_{F})_{\nu,b,\alpha} \right].
\end{equation}

Note that Eq.~(\ref{eq:data_form_index}) does not include noise. We expect the noise term $\bm{v_n}$ to propagate through the foreground subtraction procedure like the signal $\bm{v_{HI}}$ and the perturbed signal $\mbf{G}\bm{v_{HI}}$, assuming that the foreground filter removes the foregrounds but preserves both the signal and noise and that the foreground terms $\bm{v_F}$ and $\mbf{G}\bm{v_F}$ in Eq.~(\ref{eq:data_form}) dominate over the signal and noise terms. The noise bias can be removed later in the power spectrum estimation step, for example, by cross-correlating data from two seasons of observation over the same sky. 

We can use a traditional linear foreground filter $\mbf{K}$ to obtain the estimated signal
\begin{equation} \label{eq:s_hat_form}
    \bm{\hat{v}_{HI}} = \mbf{K}\bm{v_d} \approx \mbf{KG}\bm{v_{F}}.    
\end{equation}
Note that although the foreground filter $\mbf{K}$ in CHIME simulations is the KL filter \citep{mmode}, the formalism works with any other type of linear filters as well. The linear filter $\mbf{K}$ cannot remove the foregrounds coupled with systematics introduced by $\mbf{G}$, so the estimated signal in Eq.~(\ref{eq:s_hat_form}) is dominated by foreground residuals $\mbf{KG}\bm{v_F}$ (as long as the foreground filter adequately removes the intrinsic foregrounds, i.e., $\mbf{K}\bm{v_F} \ll \bm{v_{HI}}$). Similarly, we can obtain the foreground estimate
\begin{equation} \label{eq:f_hat_form}
    \bm{\hat{v}_F} = \mbf{A}\bm{v_d} \approx \bm{v_F},
\end{equation}
where $\mbf{A}$ can be any linear filter that gives a foreground estimate. For example, we can choose $\mbf{A} = \mbf{I} - \mbf{K}$, which means using the signal-estimate-subtracted data as the foreground estimate. In this study, we simply use $\mbf{A} = \mbf{I}$ since the data visibilities in the CHIME simulation are already a good estimate of the foregrounds (because the foregrounds are much brighter than the signal).

By the formalism derived in \citep{haochen_first_paper}, the telescope systematics $g_i$ can be estimated via cross-correlating the estimated signal and foregrounds
\begin{equation} \label{eq:y_hat_form}
    \hat{y}_i = \frac{\bm{\hat{v}_F}^\dagger\mbf{E_i}^\dagger\bm{\hat{v}_{HI}}}{\bm{\hat{v}_F}^\dagger\mbf{D_i}\bm{\hat{v}_F}},
\end{equation}
where $\mbf{E_i}$ is a matrix that determines over what subspace the estimated signal and foregrounds should be cross-correlated, and $\mbf{D_i}$ entails how the normalization is done. Previous work \citep{haochen_first_paper} has shown that having $\mbf{E_i} \propto \mbf{\Gamma_i}$ and $\mbf{D_i} = \mbf{E_i}^\dagger \mbf{E_i}$ are generally good choices. For example, to estimate the bandpass error $g_\nu$, with $\hat{y}_i \rightarrow \hat{y}_\nu$, $\mbf{E_i} \rightarrow \mbf{E_\nu}$, and $\mbf{D_i} \rightarrow \mbf{D_\nu}$ in Eq.~(\ref{eq:y_hat_form}), we can choose
\begin{align}
    \mbf{E_\nu} & = \mbf{K} \mbf{\Gamma_\nu} \label{eq:E_nu} \\
    \mbf{D_\nu} & = \mbf{E_\nu}^\dagger \mbf{E_\nu} \label{eq:D_nu}.
\end{align}
The reason for including $\mbf{K}$ in Eq.~(\ref{eq:E_nu}) is to project the estimated foregrounds $\bm{\hat{v}_F}$ to the KL space before cross-correlating it with the estimated signal $\bm{\hat{v}_{HI}}$. This is necessary since in Eq.~(\ref{eq:s_hat_form}), $\bm{\hat{v}_{HI}}$ has already been projected to the KL space by filter $\mathbf{K}$. This postfiltered KL space is different from the original space of the data visibilities and estimated foregrounds $\bm{\hat{v}_F}$.

The estimated systematics vector $\hat{y}_i$ in Eq.~(\ref{eq:y_hat_form}) is a linear combination of the true systematics $g_i$. This can be easily seen by plugging Eq.~(\ref{eq:G_form}) into (\ref{eq:s_hat_form}) and then (\ref{eq:s_hat_form}) into~(\ref{eq:y_hat_form}) , giving
\begin{equation} \label{eq:window_form}
    \hat{y}_i = \sum_{i^\prime} \frac{\bm{\hat{v}_F}^\dagger\mbf{E_i}^\dagger\mbf{K}\mbf{\Gamma_{i^\prime}}\bm{v_F}}{\bm{\hat{v}_F}^\dagger\mbf{D_i}\bm{\hat{v}_F}} g_{i^\prime} = \sum_{i^\prime} \mbf{W}_{i i^\prime}g_{i^\prime},
\end{equation}
where
\begin{equation} \label{eq:window_def_form}
    \mbf{W}_{i i^\prime} = \frac{\bm{\hat{v}_F}^\dagger\mbf{E_i}^\dagger\mbf{K}\mbf{\Gamma_{i^\prime}}\bm{v_F}}{\bm{\hat{v}_F}^\dagger\mbf{D_i}\bm{\hat{v}_F}}
\end{equation}
is the window matrix. Equation~(\ref{eq:window_def_form}) contains the true foregrounds $\bm{v_F}$ which we do not know. At the expense of introducing second order errors in the perturbations ($\propto \mbf{G}^2 \bm{v_F}$), we can compute Eq.~(\ref{eq:window_def_form}) using the estimated foregrounds $\bm{\hat{v}_F}$ in place of the true foregrounds $\bm{v_F}$, namely
\begin{equation} \label{eq:window_approx_form}
    \mbf{W}_{i i^\prime} \approx \frac{\bm{\hat{v}_F}^\dagger\mbf{E_i}^\dagger\mbf{K}\mbf{\Gamma_{i^\prime}}\bm{\hat{v}_F}}{\bm{\hat{v}_F}^\dagger\mbf{D_i}\bm{\hat{v}_F}}.
\end{equation}

To obtain the unwindowed systematics estimates $\hat{g}_i$, we multiply the pseudoinverse of the window matrix $\mbf{W}^+$ to the estimated systematics
\begin{equation} \label{eq:g_hat_form}
    \hat{g}_i = \sum_{i^\prime} \mbf{W}^+_{i i^\prime}\hat{y}_{i^\prime}.
\end{equation} 
The reason for using pseudoinverse is that the window matrix will not be full rank. This is because the KL filter has been applied to the data, which has removed the frequency modes occupied by intrinsic foregrounds. Real experimental data also have RFI flagging that masks contaminated frequency channels. The pseudoinverted window matrix aims to recover the frequency modes that remain in the data.

With the systematics estimated, we can reconstruct the perturbation matrix from Eq.~(\ref{eq:G_form})
\begin{equation} \label{eq:matrix_G_hat_form}
    \mbf{\hat{G}} = \sum_i \hat{g}_i \mbf{\Gamma_i}
\end{equation}
and subtract foreground residuals from the estimated signal to obtain the cleaned signal
\begin{equation} \label{eq:s_tilde_form}
    \bm{\tilde{v}_{HI}} = \bm{\hat{v}_{HI}} - \mbf{K \hat{G}}\bm{\hat{v}_F}. 
\end{equation}
For example, to clean visibilities perturbed by bandpass gains, we just need to carry through Eqs.~(\ref{eq:y_hat_form}) to (\ref{eq:s_tilde_form}) with indices $i \rightarrow \nu$ and $i^\prime \rightarrow \nu^\prime$ and use $\mbf{\Gamma_\nu}$, $\mbf{E_\nu}$, and $\mbf{D_\nu}$ defined in Eqs.~(\ref{eq:gamma_nu_form}), (\ref{eq:E_nu}), and (\ref{eq:D_nu}), respectively.

Note that because of the approximation we used in Eq.~(\ref{eq:window_approx_form}), the cleaned signal is still contaminated by foregrounds coupled with second-order perturbations, namely
\begin{equation}
    \bm{\tilde{v}_{HI}} \sim \bm{v_{HI}} + \mbf{G}^2 \bm{v_F}.
\end{equation}
In observational measurements, gains are estimated through a calibration process, and therefore the systematics represent calibration errors and uncharacterized instrumental effects. These are usually small (at the percent or subpercent level).  As long as the second-order errors are not brighter than the signal, namely $\mbf{G}^2 \bm{v_F} < \bm{v_{HI}}$, HyFoReS can sufficiently remove residual foreground contamination and recover the true signal.

Note that HyFoReS is similar to traditional self-calibration \citep{Wijnholds_2009} in the sense that both aim to constrain and reduce systematic errors from the data, but HyFoReS is innovatively different in two ways. First, HyFoReS does not need a specific sky model but only the foreground and signal covariance as required by the linear foreground filter. This reduces the risks of inaccurate sky models. Second, HyFoReS uses the 21-cm specific foreground-signal hierarchy, namely the signal being subdominant to foregrounds, and assumes that the telescope systematics are small. These conditions allow us to derive a first-order estimate of systematic errors analytically from the data rather than from parameter-inference techniques such as MCMC. Although not explicitly shown here, the analytical approach provides us with a means to control the non-linearity of the method and estimate signal loss through a perturbative expansion in terms of the signal-foreground ratio.

\subsection{Mitigating complex antenna errors with HyFoReS} \label{subsec:app}

Unlike bandpass perturbations, complex antenna gain errors are complex numbers that vary from feed to feed at each frequency. The amplitude of this error can arise from the amplifier of each antenna that boosts the signal at a slightly different rate, while the phase can arise from relative cable length differences between the antennas, introducing a delay error among the signals received from the feeds. In addition, we assume that the gain errors are time independent throughout this work.

To mitigate antenna-dependent gain errors in CHIME simulations, we need to consider two additional complexities that were not discussed in the previous section or in earlier work, such as \citep{haochen_first_paper}. First, each CHIME antenna has two polarizations: $X$ along the east-west direction and $Y$ along the north-south direction. Complex antenna errors can vary between the two polarizations. 

The second complexity is that the KL foreground filter $\mbf{K}$ in CHIME simulations is linear only on real numbers but nonlinear on complex numbers. In detail, when given visibilities $\bm{v_d}$ and a real parameter $r$, we have $\mbf{K}r\bm{v_d} = r\mbf{K}\bm{v_d}$. However, if $c$ is a complex parameter, then $\mbf{K}c\bm{v_d} \neq c\mbf{K}\bm{v_d}$. This is because in CHIME simulations, the visibilities $\bm{v_d}$ contain only half of the full visibilities $\{(v_d)_{ij}|i < j\}$ due to the conjugation symmetry $(v_d)^*_{ij} = (v_d)_{ji}$, where $(v_d)_{ij}$ represents the visibility formed from the $i$th and $j$th antennas. Although redundant, both $(v_d)_{ij}$ and $(v_d)^*_{ij}$ are physical modes of the sky measurement, so the foreground filter $\mbf{K}$ may need to use a combination of visibilities and complex-conjugated visibilities from $\bm{v_d}$ to eliminate intrinsic foregrounds. In other words, the operator $\mbf{K}$ is effectively the sum of two operators, one acting on the half visibilities $\bm{v_d}$ and the other acting on their complex conjugates, namely
\begin{equation}
\label{eq:Kc1}
    \mbf{K}c\bm{v_d} = \mbf{K_1}c\bm{v_d} + \mbf{K_2}c^*\bm{v_d}^*,
\end{equation}
where $\mbf{K_1}$ and $\mbf{K_2}$ are linear operators. Although not affecting any real parameter, the operation of complex conjugation introduces a minus sign to the imaginary component of $c$, that is, the imaginary unit $i$ does not commute with the filter $\mbf{K}$. To mitigate this issue, we write any complex parameter as $c = g + ih$, where $g$ and $h$ are real numbers that can propagate through the pipeline linearly as
\begin{equation}
\label{eq:Kc2}
    \mbf{K}c\bm{v_d} = \mbf{K}g\bm{v_d} + \mbf{K}ih\bm{v_d} = g\mbf{K}\bm{v_d} + h\mbf{K}i\bm{v_d},
\end{equation}
where to get the first equal sign, we have used the fact that even though $\mbf{K}$ is nonlinear, it is still distributive. Also note that in our simulations, the real part and imaginary parts have comparable magnitude.

We now apply the general formalism in Sec.~\ref{subsec:form_rev} considering these complexities. Let $({v_d})_{\nu, ipjp^\prime, \alpha}$ represent the observed visibilities made by polarization $p$ of the $i$th antenna and polarization $p^\prime$ of the $j$th antenna ($p$, $p^\prime$ = $X$ or $Y$) at frequency $\nu$ and RA $\alpha$. Following the form of Eq.~(\ref{eq:data_form_index}), the observed visibilities in this case are
\begin{equation} \label{eq: v_d_1}
\begin{split}
    ({v_d})_{\nu, ipjp^\prime, \alpha} = & (1 + c_{\nu, i p})(1 + c^*_{\nu, j p^\prime}) \\
    & [({v_{HI}})_{\nu, ipjp^\prime, \alpha} + ({v_{F}})_{\nu, ipjp^\prime, \alpha} ],
\end{split}
\end{equation}
where $v_{HI}$ and $v_{F}$ are the true HI and foreground visibilities as before, and $c_{\nu, i p}$ is the gain error from polarization $p$ of the $i$th antenna at frequency $\nu$ ($c_{\nu, j p^\prime}$ defined accordingly). The simulated telescope has many redundant baselines that measure the same sky information despite different gain errors. We can average visibilities from redundant baselines to form stacked visibilities as
\begin{equation} \label{eq: v_d_2}
\begin{split}
    (v_d)_{\nu, bx,\alpha} = \frac{1}{N_{ipjp^\prime}}\sum_{(i,j) \in b} \sum_{(p,p^\prime) \in x} v_{\nu, ipjp^\prime, \alpha},
\end{split}
\end{equation}
where $b$ represents stacked baselines, $x = XX$, $XY$, $YX$, or $YY$ stands for visibilities' polarizations, and $N_{ipjp^\prime}$ counts the redundancy. Plugging Eq.~(\ref{eq: v_d_1}) into Eq.~(\ref{eq: v_d_2}), we have
\begin{equation} \label{eq: v_d_3}
    (v_d)_{\nu, bx,\alpha} = (1 + g_{\nu, bx} + i h_{\nu, bx}) [(v_{HI})_{\nu, bx, \alpha} + (v_F)_{\nu, bx, \alpha}],
\end{equation}
where
\begin{equation} \label{eq:stack_error}
    1 + g_{\nu, bx} + i h_{\nu, bx} = \frac{1}{N_{ipjp^\prime}}\sum_{(i,j) \in b} \sum_{(p,p^\prime) \in x} (1 + c_{\nu, i p})(1 + c^*_{\nu, j p^\prime}).
\end{equation}
Instead of estimating the antenna-dependant gains $c_{\nu, i p}$, we will estimate the errors on the stacked visibilities, $g_{\nu, bx}$ and $h_{\nu, bx}$. This is because the foreground filter $\mbf{K}$ operates on the stacked visibilities in the simulation, and there are fewer parameters to estimate in this space.

We observe that Eq.~(\ref{eq: v_d_3}) now matches the form of Eq.~(\ref{eq:data_form}) by specifying the perturbation matrix as
\begin{equation} \label{eq:G_sim}
    \mbf{G} = \sum_{\nu,bx} (g_{\nu,bx} + i h_{\nu,bx}) \mbf{\Gamma_{\nu,bx}},
\end{equation}
with
\begin{equation} \label{eq:Gamma_sim}
\begin{split}
       & (\Gamma_{\nu,bx})_{(\nu^\prime,b^\prime x^\prime, \alpha^\prime)(\nu^{\dprime},b^{\dprime} x^{\dprime}, \alpha^{\dprime})} \\ & = I_{(\nu^\prime,b^\prime x^\prime, \alpha^\prime)(\nu^{\dprime},b^{\dprime} x^{\dprime}, \alpha^{\dprime})} \delta_{\nu \nu^\prime} \delta_{b b^\prime} \delta_{x x^\prime},
\end{split}  
\end{equation}
where $I_{(\nu^\prime,b^\prime x^\prime, \alpha^\prime)(\nu^{\dprime},b^{\dprime} x^{\dprime}, \alpha^{\dprime})} = \delta_{\nu^\prime \nu^{\dprime}} \delta_{b^\prime b^{\dprime}} \delta_{x^\prime x^{\dprime}} \delta_{\alpha^\prime \alpha^{\dprime}}$ is the identity matrix. Equations~(\ref{eq:G_sim}) and (\ref{eq:Gamma_sim}) state that each complex error $(g_{\nu,bx} + i h_{\nu,bx})$ affects visibilities with beam $b$, polarization $x$, and frequency $\nu$ in all RA (the stationarity over RA comes from our assumption that the gain errors are time independent). 

Note that from Eq.~(\ref{eq:G_sim}), we can define two separate derivative matrices for the real and imaginary parts of the gain perturbations, namely
\begin{equation} \label{eq:G_two_m}
    \mbf{G} = \sum_{\nu,bx} g_{\nu,bx}\mbf{\Gamma_{\nu,bx}} +  h_{\nu,bx}\mbf{\Delta_{\nu,bx}},
\end{equation}
where 
\begin{equation}
    \mbf{\Delta_{\nu,bx}} = i \mbf{\Gamma_{\nu,bx}},
\end{equation}
is the derivative matrix of the imaginary part of the gain perturbations.

Now we apply the foreground filter $\mbf{K}$ on the stacked visibilities, similar to Eq.~(\ref{eq:s_hat_form})
\begin{equation} \label{eq:v_HI_hat}
\begin{split}
    \bm{\hat{v}_{HI}} & \approx \mbf{K G} \bm{v_F} \\
    & = \sum_{\nu, bx} g_{\nu, bx}(\mbf{K \Gamma_{\nu, bx}} \bm{v_F}) + h_{\nu, bx}(\mbf{K \Delta_{\nu, bx}} \bm{v_F}),
\end{split}
\end{equation}
where we arrived at the second line using Eq.~(\ref{eq:G_two_m}). To get a foreground estimate, we use the fact that the 21-cm radio sky is dominated by foregrounds (since the foreground to HI signal ratio is about $10^5$), so the observed visibilities are already a good measurement of the foregrounds. We therefore adopt
\begin{equation} \label{eq:v_F}
    \bm{\hat{v}_F} = \bm{v_d}.
\end{equation}
Note that this is equivalent to taking $\mbf{A} = \mbf{I}$ in Eq.~(\ref{eq:f_hat_form}).

Now we can cross-correlate the estimated foregrounds with the estimated signal to estimate the gain perturbations, similar to Eq.~(\ref{eq:y_hat_form}). However, this is now more complicated, since we treat the real and imaginary parts of the gain perturbations as independent parameters. As a result, we will need two types of cross-correlation matrices ($\mbf{E_i}$ in Eq.[\ref{eq:y_hat_form}]): $\mbf{E^R_{\nu,bx}}$ for the real parts and $\mbf{E^I_{\nu,bx}}$ for the imaginary parts. In Sec.~\ref{subsec:form_rev}, we mentioned that having $\mbf{E_i} \propto \mbf{\Gamma_i}$ is a good choice, so it is natural to pick $\mbf{E^R_{\nu,bx}} \propto \mbf{\Gamma_{\nu,bx}}$ and $\mbf{E^I_{\nu,bx}} \propto \mbf{\Delta_{\nu,bx}}$. Similar to the bandpass case (Eq.~[\ref{eq:E_nu}]), we choose the cross-correlation matrices as
\begin{align}
    & \mbf{E^R_{\nu,bx}} = \mbf{K} \mbf{\Gamma_{\nu,bx}} \label{eq:E_R} \\
    & \mbf{E^I_{\nu,bx}} = \mbf{K} \mbf{\Delta_{\nu,bx}} \label{eq:E_I},
\end{align}
and their corresponding normalization matrices as
\begin{align}
    & \mbf{D^R_{\nu,bx}} = (\mbf{E^R_{\nu,bx}})^\dagger (\mbf{E^R_{\nu,bx}}) \\
    & \mbf{D^I_{\nu,bx}} = (\mbf{E^I_{\nu,bx}})^\dagger (\mbf{E^I_{\nu,bx}}).
\end{align}

Like Eq.~(\ref{eq:y_hat_form}), we can estimate the real and imaginary parts of the gain perturbations as
\begin{equation} \label{eq:g_hat_sim}
    \bm{\hat{g}}_{\nu, bx} = \frac{\bm{\hat{v}^\dagger_F}(\mbf{E^R_{\nu,bx}})^\dagger\bm{\hat{v}_{HI}}}{\bm{\hat{v}^\dagger_F}\mbf{D^R_{\nu,bx}}\bm{\hat{v}_F}},
\end{equation}
and 
\begin{equation} \label{eq:h_hat_sim}
    \bm{\hat{h}}_{\nu, bx} = \frac{\bm{\hat{v}^\dagger_F}(\mbf{E^I_{\nu,bx}})^\dagger\bm{\hat{v}_{HI}}}{\bm{\hat{v}^\dagger_F}\mbf{D^I_{\nu,bx}}\bm{\hat{v}_F}},
\end{equation}
respectively.

To obtain the window matrix, we plug Eq.~(\ref{eq:v_HI_hat}) into Eqs.~(\ref{eq:g_hat_sim}) and (\ref{eq:h_hat_sim}) (similar to how we arrived at Eq.~[\ref{eq:window_def_form}] from Eq.~[\ref{eq:window_form}]). Notice that $\bm{\hat{v}_{HI}}$ in Eq.~(\ref{eq:v_HI_hat}) has contributions from both the real and imaginary parts of the gain perturbations. This means that the window matrix mixes the real and imaginary parts. It is then convenient to pack the real and imaginary parts $\bm{\hat{g}}$ and $\bm{\hat{h}}$ into one vector array, and rewrite Eqs.~(\ref{eq:g_hat_sim}) and (\ref{eq:h_hat_sim}) as
\begin{equation} \label{eq:g_hat_vec}
    \begin{bmatrix}
        \bm{\hat{g}} \\ \bm{\hat{h}}
    \end{bmatrix}
    =
    \begin{bmatrix}
        \mbf{W^{RR}} & \mbf{W^{RI}} \\ \mbf{W^{IR}} & \mbf{W^{II}}
    \end{bmatrix}
    \begin{bmatrix}
        \bm{{g}} \\ \bm{{h}}
    \end{bmatrix},    
\end{equation}
where
\begin{align}
    & (\mbf{W^{RR}})_{(\nu,bx)(\nu^\prime,b^\prime x^\prime)} = \frac{\bm{\hat{v}^\dagger_F}(\mbf{E^R_{\nu,bx}})^\dagger \mbf{K \Gamma_{\nu^\prime,b^\prime x^\prime}}\bm{\hat{v}_F}}{\bm{\hat{v}^\dagger_F}\mbf{D^R_{\nu,bx}}\bm{\hat{v}_F}} \label{eq:W_RR} \\
    & (\mbf{W^{RI}})_{(\nu,bx)(\nu^\prime,b^\prime x^\prime)} = \frac{\bm{\hat{v}^\dagger_F}(\mbf{E^R_{\nu,bx}})^\dagger \mbf{K \Delta_{\nu^\prime,b^\prime x^\prime}}\bm{\hat{v}_F}}{\bm{\hat{v}^\dagger_F}\mbf{D^R_{\nu,bx}}\bm{\hat{v}_F}} \\
    & (\mbf{W^{IR}})_{(\nu,bx)(\nu^\prime,b^\prime x^\prime)} = \frac{\bm{\hat{v}^\dagger_F}(\mbf{E^I_{\nu,bx}})^\dagger \mbf{K \Gamma_{\nu^\prime,b^\prime x^\prime}}\bm{\hat{v}_F}}{\bm{\hat{v}^\dagger_F}\mbf{D^I_{\nu,bx}}\bm{\hat{v}_F}} \\
    & (\mbf{W^{II}})_{(\nu,bx)(\nu^\prime,b^\prime x^\prime)} = \frac{\bm{\hat{v}^\dagger_F}(\mbf{E^I_{\nu,bx}})^\dagger \mbf{K \Delta_{\nu^\prime,b^\prime x^\prime}}\bm{\hat{v}_F}}{\bm{\hat{v}^\dagger_F}\mbf{D^I_{\nu,bx}}\bm{\hat{v}_F}} \label{eq:W_II}
\end{align}
are block matrices that make up the full window matrix.

To obtain the unwindowed gain perturbation estimates, similar to Eq.~(\ref{eq:g_hat_form}), we perform a pseudoinverse of the window and apply it to the estimated gains
\begin{equation} \label{eq:g_tilde_com}
        \begin{bmatrix}
        \bm{\tilde{g}} \\ \bm{\tilde{h}}
    \end{bmatrix}
    =
      \left(  \begin{bmatrix}
        \mbf{W^{RR}} & \mbf{W^{RI}} \\ \mbf{W^{IR}} & \mbf{W^{II}}
    \end{bmatrix} \right)^+
    \begin{bmatrix}
        \bm{\hat{g}} \\ \bm{\hat{h}}
    \end{bmatrix},
\end{equation}
where $\bm{\tilde{g}}$ and $\bm{\tilde{h}}$ are the unwindowed estimate for the real and imaginary parts of the gain perturbations, respectively.

There are two subtleties. First, note that the parameters we want to recover, namely $\bm{g}$ and $\bm{h}$ on the right-hand side of Eq.~(\ref{eq:g_hat_vec}), are real numbers, but the estimates $\bm{\hat{g}}$ and $\bm{\hat{h}}$ on the left-hand side, which we obtained from Eqs.~(\ref{eq:g_hat_sim}) and (\ref{eq:h_hat_sim}), are complex in general due to noise in our estimation. 
This is equivalent to say that the window matrix in Eq.~(\ref{eq:g_hat_vec}) has an imaginary part due to noise, but we are only interested in the real part, namely
\begin{equation} \label{eq:g_hat_vec_real}
\begin{split}
    \mbf{Re} \begin{bmatrix}
        \bm{\hat{g}} \\ \bm{\hat{h}}
    \end{bmatrix}
    & = \left ( \mbf{Re}
    \begin{bmatrix}
        \mbf{W^{RR}} & \mbf{W^{RI}} \\ \mbf{W^{IR}} & \mbf{W^{II}}
    \end{bmatrix} \right )
    \begin{bmatrix}
        \bm{{g}} \\ \bm{{h}}
    \end{bmatrix}.    \\
    & 
\end{split}
\end{equation}
To ensure the unwindowed estimates $\bm{\tilde{g}}$ and $\bm{\tilde{h}}$ are real in Eq.~(\ref{eq:g_tilde_com}), we pseudoinvert the real part of the window matrix in Eq.~(\ref{eq:g_hat_vec_real}) and act it on the real part of the vector $\bm{\hat{g}}$ and $\bm{\hat{h}}$.

The second subtlety is that in our data visibilities, the co-polarization visibilities ($x = XX $ or $YY$) are much better measured than the cross-polarization visibilities ($x = XY $ or $YX$). This is because the foreground contribution to cross-polarization is much weaker than the foreground contribution to co-polarization. As a result, the cross-polarization visibilities have much lower intensities and are much more sensitive to noise. This, in turn, implies that the gain perturbation estimates $\hat{g}_{\nu, bx}$ and $\hat{h}_{\nu, bx}$ have much greater uncertainties in the case of $x = XY $ or $YX$. These uncertainties will eventually threaten our measurement for the gains of the co-polarization visibilities through the window mixing. The solution is simple. Since the cross-polarization visibilities are much  smaller in intensities,
\begin{align}
    & (v_{F})_{\nu, bx, \alpha} |_{x = XY,YX} \ll (v_{F})_{\nu, bx, \alpha} |_{x = XX, YY} \\
    & (v_{HI})_{\nu, bx, \alpha} |_{x = XY,YX} \ll (v_{HI})_{\nu, bx, \alpha} |_{x = XX, YY},
\end{align}
we can regard the effect of the gain perturbations in cross-polarizations as the second-order errors and ignore them in parameter estimation and foreground subtraction.


Now we update Eq.~(\ref{eq:g_tilde_com}) by considering these two subtleties mentioned above and obtain the unwindowed gain parameters as
\begin{equation} \label{eq:g_tilde_real}
        \begin{bmatrix}
        \bm{\tilde{g}} \\ \bm{\tilde{h}}
    \end{bmatrix}
    =
      \left(  \mbf{Re} \begin{bmatrix}
        \mbf{W^{RR}} & \mbf{W^{RI}} \\ \mbf{W^{IR}} & \mbf{W^{II}}
    \end{bmatrix} \right)^+_{x = XX, YY}
    \left ( \mbf{Re}
    \begin{bmatrix}
        \bm{\hat{g}} \\ \bm{\hat{h}}
    \end{bmatrix} \right ),
\end{equation}
where the restriction $x = XX, YY$ means we only keep elements in the pseudoinverted window matrix associated with the co-polarizations. In other words, we pseudoinvert the real part of the full window matrix first and then set all rows and columns corresponding to $x = XY, YX$ to zero.

With $\bm{\tilde{g}}$ and $\bm{\tilde{h}}$ from Eq.~(\ref{eq:g_tilde_real}), we can now reconstruct the perturbation matrix
\begin{equation} \label{eq:G_hat_sim}
    \mbf{\hat{G}} = \sum_{\nu,bx} \tilde{g}_{\nu,bx}\mbf{\Gamma_{\nu,bx}} +  \tilde{h}_{\nu,bx}\mbf{\Delta_{\nu,bx}},
\end{equation}
and subtract foreground residuals (at the first order in gain perturbations) from the estimated signal $\bm{\hat{v}_{HI}}$ and obtain the cleaned signal 
\begin{equation} \label{eq:s_tilde_sim}
\begin{split}
        \bm{\tilde{v}_{HI}} & = \bm{\hat{v}_{HI}} - \mbf{K \hat{G}} \bm{\hat{v}_F},
\end{split}
\end{equation}
similar to Eq.~(\ref{eq:s_tilde_form}) in the bandpass case.


\section{simulation pipeline} \label{sec:sim}

We first introduce the default CHIME simulation pipeline in Sec.~\ref{subsec:default}, followed by the additional steps that we add for this study in Sec.~\ref{subsec:modif}.

\subsection{CHIME simulation pipeline} \label{subsec:default}

\begin{figure}
    \centering
    \includegraphics[width=0.8\linewidth]{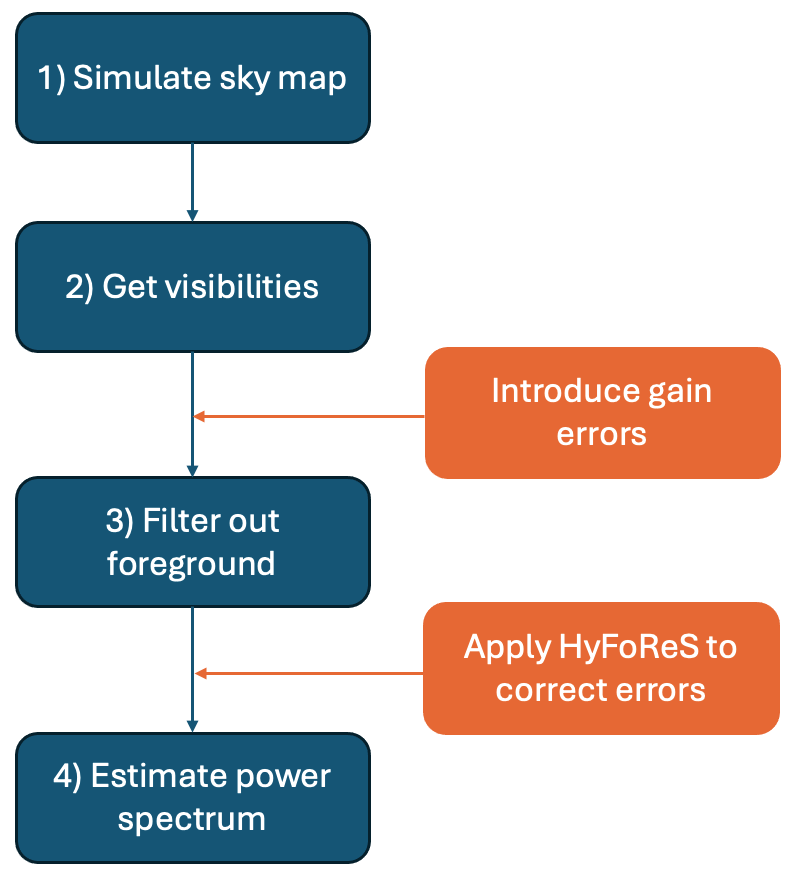}
    \caption{Flow chart showing the key steps in the default CHIME simulation pipeline (dark blue boxes) and the additional steps added for this study (orange boxes). In the default pipeline, we simulate a sky map made from 21 cm signal and foregrounds, generate visibilities observed by a telescope given the map, filter out the foreground components of the visibilites, and finally estimate the 2-dimensional HI power spectrum. For the purpose of this study, we introduce gain errors to the visibilities after step 2, and then apply HyFoReS to correct for gain errors after step 3.}
    \label{fig:flow_chart}
\end{figure}

The CHIME simulations were first introduced in \citep{first_mmode, mmode}. Part of the simulations was also used in the CHIME stacking analysis \citep{chime_stacking} and in its related works \citep{chime_lyman, wang2024}. The simulation pipeline is based on modules from the \texttt{Radio Cosmology} packages\footnote{https://github.com/radiocosmology}, with a brief overview shown in Fig.~\ref{fig:flow_chart} and summarized below. 

The first step in the simulation is to generate the sky maps. This is done using \texttt{cora}, which produces \texttt{HEALPix} maps for the HI signal and foregrounds. The signal map is made by drawing Gaussian realizations from a theoretic power spectrum of 21-cm brightness temperature (see the appendix of \citep{mmode} for details). The foreground map, in principle, has two components: 1) galactic synchrotron emission extrapolated from the Haslam map \citep{haslam} but with randomly generated frequency and angular fluctuations added in, and 2) point sources derived from real surveys, a synthetic population of dimmer point sources, and a Gaussian random field to model an unresolved background of sources. However, in the current stage of pipeline development, the KL foreground filter does not include a covariance of point sources and can only remove the intrinsic Galactic foreground. Ongoing work will expand the foreground model for the KL filter in the future. Due to this technical limitation, we turn off the point sources and only use the Galactic synchrotron emission as the foreground model for this study.

The next step is to simulate a telescope. This is done with \texttt{driftscan} by computing the beam transfer matrices (see \citep{mmode} for a review), which are linear operations that transform the sky map to visibilities for a given telescope setup. In this study, we use two setups. Both are modeled after the CHIME pathfinder \citep{pathfinder}, consisting of two 20 meter by 20 meter cylinders, observing the 400 MHz to 500 MHz band with a system temperature of 50 K. The two setups differ in the number of frequency channels, number of antenna feeds per cylinder, and integration time (see Table~\ref{tab:my_label}). 

The pipeline then uses the KL transform to filter out foregrounds in the simulated visibilities (see \citep{first_mmode, mmode} for a review of the KL filter). Finally, a quadratic estimator is used to measure the 2-dimensional HI power spectrum from the filtered visibilities. Note that no systematics are simulated in the default CHIME pipeline.

\begin{table*}[]
    \centering
    \begin{tabular}{|c|c|c|c|c|}
     \hline
     Setting & Freq. channels (400 to 500 MHz) & Feeds per cylinder & Feed spacing (meters) & Integration time (days) \\
     \hline
     Full-pathfinder & 40 & 64 & 0.3 & 730 \\
     Half-pathfinder & 20 & 32 & 0.6 & 18250 \\
     \hline
    \end{tabular}
    \caption{Comparison of the two telescope settings. Both telescopes are made with two 20-meter-by-20-meter cylinders and observe in the 400 MHz to 500 MHz band, but differ in the number of frequency channels and antenna feeds. In addition, the half pathfinder has a longer integration time to compensate for its lower sensitivity compared with the full pathfinder.}
    \label{tab:my_label}
\end{table*}

\subsection{Implementing HyFoReS to mitigate gain perturbations} \label{subsec:modif}

We then introduce gain-type errors to the simulation and subsequently use HyFoReS to subtract the gain-induced foreground contamination from the filtered visibilities. We do so in two separate cases: bandpass perturbations on the full-pathfinder setup and complex antenna perturbations on the half-pathfinder telescope. This is to reduce the computational cost with complex antenna perturbations.

To simulate bandpass errors, we first obtain the simulated visibilities as the sum of the unperturbed signal and foregrounds. We then perturb the visibilities using Eqs.~(\ref{eq:data_form}) and (\ref{eq:G_form}) with $g_i \rightarrow g_\nu$, and $\Gamma_i \rightarrow \Gamma_\nu$ as defined by Eq.~(\ref{eq:gamma_nu_form}). We draw the bandpass errors $g_\nu$ from a normal distribution with an rms value of $10^{-5}$, and then repeat the process three more times, each time increasing the rms value of the errors by a factor of 10, up to the level of $10^{-2}$. This produces 4 sets of perturbed visibilities, which we will use to explore the effect of bandpass perturbations across 4 orders of magnitude and test to what extent HyFoReS can mitigate these errors. Note that we stop at the $10^{-2}$ level because antenna gains are calibrated to percent level precision or better in the real CHIME data \citep{chime_stacking}.

To simulate complex antenna perturbations, we draw the real and imaginary parts of the gains associated with each antenna $i$, polarization $p$, and frequency $\nu$ ($\text{Re}[c_{\nu, ip}]$ and $\text{Im}[c_{\nu, ip}]$ in Eq.~[\ref{eq: v_d_1}]) separately from a normal distribution with an rms value of $10^{-5}$. Note that the simulation pipeline stacks visibilities over redundant baselines in step 2 of Fig.~\ref{fig:flow_chart}, so we stack antenna gain errors $c_{\nu, ip}$ accordingly using Eq.~(\ref{eq:stack_error}) and compute the stacked visibility errors ($g_{\nu,bx}$ and $h_{\nu,bx}$) at each frequency, baseline, and polarization. We then make the perturbation matrices defined by Eqs.~(\ref{eq:G_sim}) and (\ref{eq:Gamma_sim}) and apply them to the unperturbed visibilities according to Eq.~(\ref{eq:data_form}). Similar to the bandpass case, we repeat this process with the rms of the antenna gain errors being $10^{-4}$, $10^{-3}$, and $10^{-2}$ in three additional runs.

A traditional linear foreground filter will no longer work with gain errors, leaving foreground residuals in the visibilities after step 3 of the pipeline. We now implement HyFoReS to remove the foreground residuals. HyFoReS requires an estimated signal and foreground. In this study, we use the simulated data $\bm{v_d}$ as the estimated foreground and use the filtered visibilities after step 3 as the estimated signal. To mitigate bandpass errors, we apply Eq.~(\ref{eq:y_hat_form}) with index $i \rightarrow \nu$, and use $\mbf{E_\nu}$ and $\mbf{D_\nu}$ defined by Eqs.~(\ref{eq:E_nu}) and (\ref{eq:D_nu}) to estimate the bandpass gains. To compensate for the window of the estimated gains, we follow Eqs.~(\ref{eq:window_approx_form}) and (\ref{eq:g_hat_form}) and finally use Eqs.~(\ref{eq:matrix_G_hat_form}) and (\ref{eq:s_tilde_form}) to subtract the foreground residuals to obtain the cleaned visibilities. 

The procedure for mitigating complex antenna errors is similar. We carry through Eqs.~(\ref{eq:E_R}) to (\ref{eq:h_hat_sim}) to estimate the real and imaginary parts of the stacked visibility errors, compensate for their windows using Eq.~(\ref{eq:g_tilde_real}) with the window matrices defined by Eqs.~(\ref{eq:W_RR}) to (\ref{eq:W_II}), and finally subtract the foreground residuals using Eqs.~(\ref{eq:G_hat_sim}) and (\ref{eq:s_tilde_sim}).

We then feed the cleaned visibilities to the power spectrum estimator (step 4 of the default CHIME simulation pipeline). For comparison, we also measure the power spectrum from the perturbed visibilities without implementing HyFoReS and from the unperturbed visibilities (without gain errors introduced). The results are shown in the following section.

\section{results} \label{sec:result}
\subsection{Bandpass gains}

\begin{figure*}[htb]
\includegraphics[width=2\columnwidth]{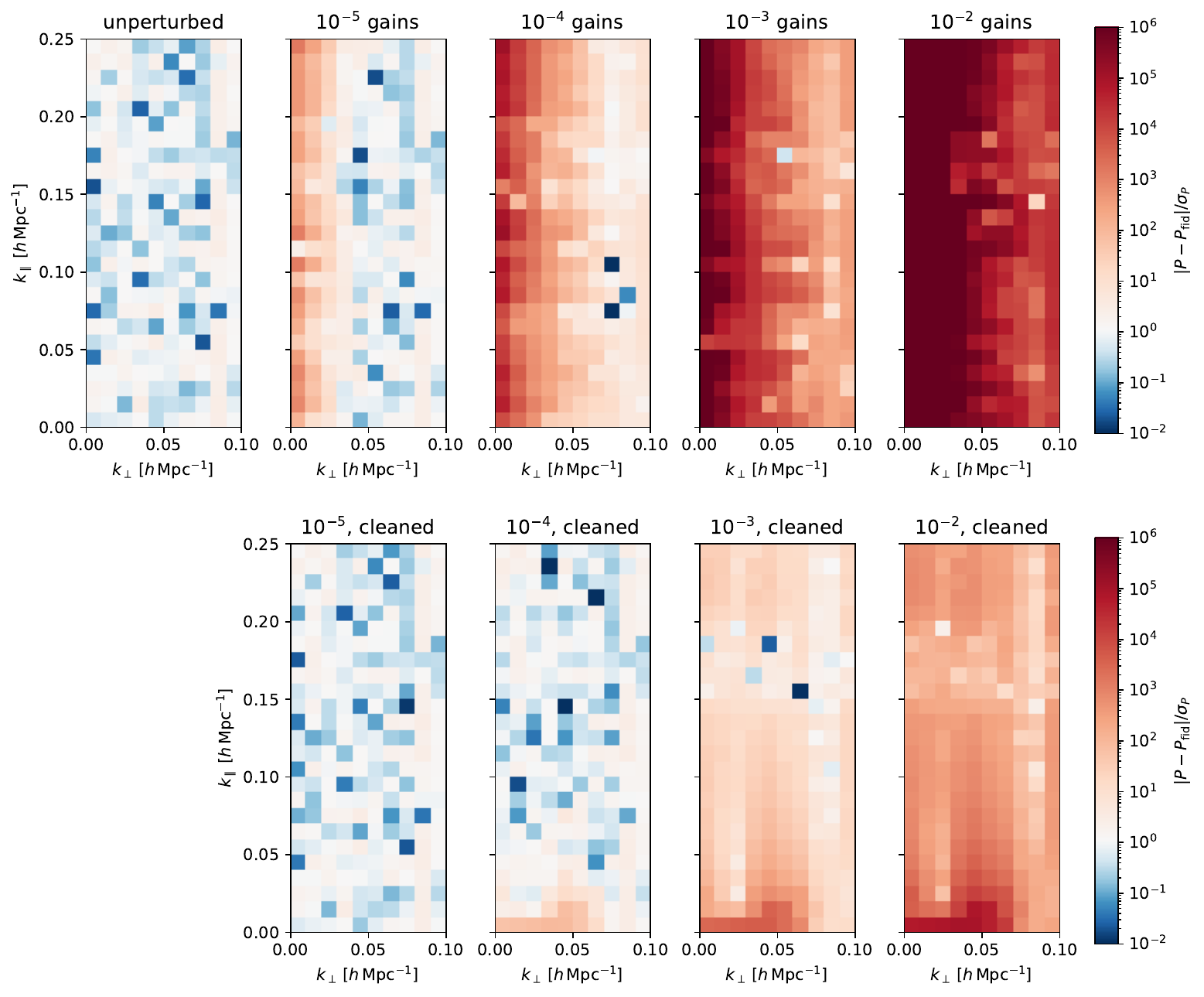}
\caption{Bias in the two-dimensional power spectra of the simulated data with the full-pathfinder telescope setting. The top panels from left to right correspond to power spectra made from the unperturbed visibilities, visibilities perturbed by $10^{-5}$, $10^{-4}$, $10^{-3}$, and $10^{-2}$ level bandpass errors, respectively, after a KL foreground filter has been applied. Due to bandpass errors, the KL filter is unable to remove all foregrounds, biasing the measurement. The bottom panels show the power spectra after HyFoReS subtracts residual foregrounds from the visibilities corresponding to the top panels. HyFoReS suppresses foreground bias to below the thermal noise for $10^{-5}$ and $10^{-4}$ level bandpass errors over most $k$ bins and reduces foreground bias by three orders of magnitude for the $10^{-3}$ and $10^{-2}$ cases.
\label{fig:BP-pert}}
\end{figure*}

In this section, we show the results of using HyFoReS to mitigate bandpass errors in the simulated data of the full pathfinder telescope. Following the procedure detailed in Sec.~\ref{subsec:modif}, we introduce bandpass errors to the visibilities and compare the power spectrum before and after HyFoReS. The results are shown in Fig.~\ref{fig:BP-pert}. Note that the plotted quantity is the bias in the measured power spectrum, i.e. the difference between the measured and true HI power spectra, divided by the one-sigma uncertainty (from thermal noise and cosmic variance) across the comoving wavenumbers along the line of sight ($k_{\parallel}$) and perpendicular to the line of sight ($k_{\perp}$). We observe that the power spectrum from the unperturbed visibilities after KL foreground filtering is consistent with the true HI power. 

After introducing bandpass errors, we observe an increasing amount of bias in the power spectrum as the rms values of bandpass perturbations increase. This is because the KL filter fails to remove foregrounds after bandpass perturbations introduce frequency-dependent multiplicative errors to the observed foregrounds. More specifically, we find foreground bias first appears at low $k_\perp$ across all $k_\parallel$ and then extends to the rest of the $k$-space as bandpass errors increase. This behavior is expected for bandpass errors coupled with Galactic foregrounds. Since $k_\parallel$ corresponds to structures along the frequency axis of the data, frequency-dependent bandpass errors introduce bias on this axis. We see the bias strongest at lower $k_\perp$, corresponding to large angular scales where the diffuse Galactic synchrotron emission dominates.


In comparison, after applying HyFoReS to KL-filtered visibilities, we find that the foreground bias has been reduced to below the noise level for the $10^{-5}$ case. Foregrounds are reduced similarly in the $10^{-4}$ case, except in the regions of low $k_\parallel$ corresponding to the ``foreground wedge,'' where the intrinsic foregrounds occupy in the $k-$space. The residual foreground bias is not completely removed for the $10^{-3}$ and $10^{-2}$ cases. When the bandpass perturbations grow large, the foreground residuals from second-order gain perturbations increase, which HyFoReS is currently unable to remove. The foreground bias is nevertheless reduced by roughly three orders of magnitude compared to that without HyFoReS. This demonstrates the effectiveness of the algorithm in mitigating bandpass perturbations.

\begin{figure*}
\includegraphics[width=2\columnwidth]{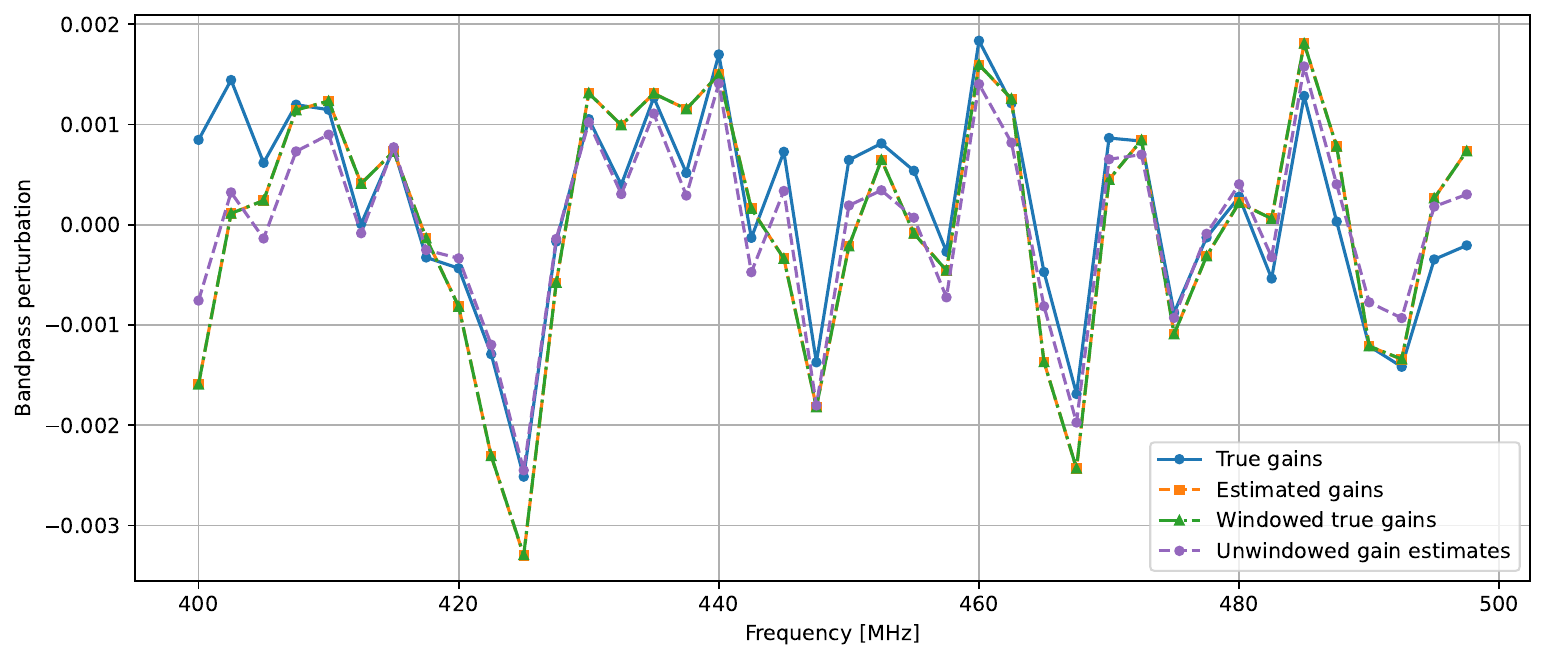}
\caption{Comparison between the true $10^{-3}$ level bandpass gains introduced to the simulation (blue), gains estimated with Eq.~(\ref{eq:y_hat_form}) (orange), true gains passed through the window matrix (green), and unwindowed gain estimates after pseudoinverting the window using Eq.~(\ref{eq:g_hat_form}) (purple). The estimated gains match the windowed true gains within $10^{-6}$, consistent with second-order gain errors. The offset between the true gains and unwindowed gain estimates is due to the frequency modes lost through the KL foreground filtering.  
\label{fig:BP-1e-3}}
\end{figure*}

To illustrate how HyFoReS estimates the bandpass gains and subtracts foreground residuals, Fig.~\ref{fig:BP-1e-3} compares the true bandpass gain errors $\bm{g}$ that we have introduced to the simulation in blue, the estimated gains by HyFoReS in orange ($\bm{\hat{y}}$ from Eq.~[\ref{eq:y_hat_form}]), the true bandpass gains multiplied with the window matrix in green ($\mbf{W}\bm{g}$, i.e., the right hand side of the second equal sign in Eq.~[\ref{eq:window_form}]), and finally the unwindowed gain estimates in purple ($\bm{\hat{g}}$ from Eq.~[\ref{eq:g_hat_form}]), for the case of $10^{-3}$ level bandpass errors. First, by Eq.~(\ref{eq:window_form}), we expect that the estimated gains $\bm{\hat{y}}$ should match the windowed true gains $\mbf{W}\bm{g}$ if our window matrix calculation is accurate and if there is no noise in the estimated gains. (Recall that Eq.~[\ref{eq:window_form}] says that the gains estimated by HyFoReS are a linear combination of the true bandpass gains, with the window matrix $\mbf{W}$ mapping the true gain perturbations to the estimated values.)

However, the second equal sign in Eq.~(\ref{eq:window_form}) is not strictly true in our implementation because we have used Eq.~(\ref{eq:window_approx_form}) to approximate the window matrix, introducing errors at the second order of gain perturbations. In addition, when computing Eq.~(\ref{eq:window_approx_form}) from Eq.~(\ref{eq:y_hat_form}), we have ignored all but the foreground components of the estimated foregrounds and signal. Other terms, such as noise in the data, can spuriously correlate with foregrounds. In Fig.~\ref{fig:BP-1e-3}, the rms deviation between the estimated gains and the windowed true gains is around 0.1\% of the true gains, accounting for the error in both the computed window matrix and noise in the gain estimation. We expect that this subpercent deviation will be coupled with the $10^{-3}$ level gains and lead to $10^{-6}$ level residual perturbations on the foreground visibilities. This illustrates second-order errors from the $10^{-3}$ level gains. 

Note the offset between the true gains and the unwindowed gain estimates in Fig.~\ref{fig:BP-1e-3}. This is because we use the pseudoinverse of the window matrix in Eq.~(\ref{eq:g_hat_form}) when recovering the true gains. Recall that the KL filter has already removed the modes corresponding to the intrinsic foregrounds before we estimate the gains, so these modes are unrecoverable. We expect that the removed modes have a smooth frequency dependence (corresponding to low-pass frequency modes) just like the intrinsic foregrounds. In Fig.~\ref{fig:BP-1e-3}, we observe that the offset between the unwindowed and true gains indeed varies smoothly with frequencies: the unwindowed gains initially lie below the true gains at 400 MHz, approach the true gains around 430 MHz, and then dip again before catching up with the true gains and finally taking over just before 500 MHz. Despite this offset, the unwindowed gains can be used to remove residual foregrounds, as demonstrated by the results shown in Fig.~\ref{fig:BP-pert}. This is because the modes lost in the gains are not present in the residual foregrounds either, so it is sufficient (and, in fact, necessary) to use the unwindowed gain estimates to subtract residual foregrounds.

\begin{figure*}[htb]
\includegraphics[width=2\columnwidth]{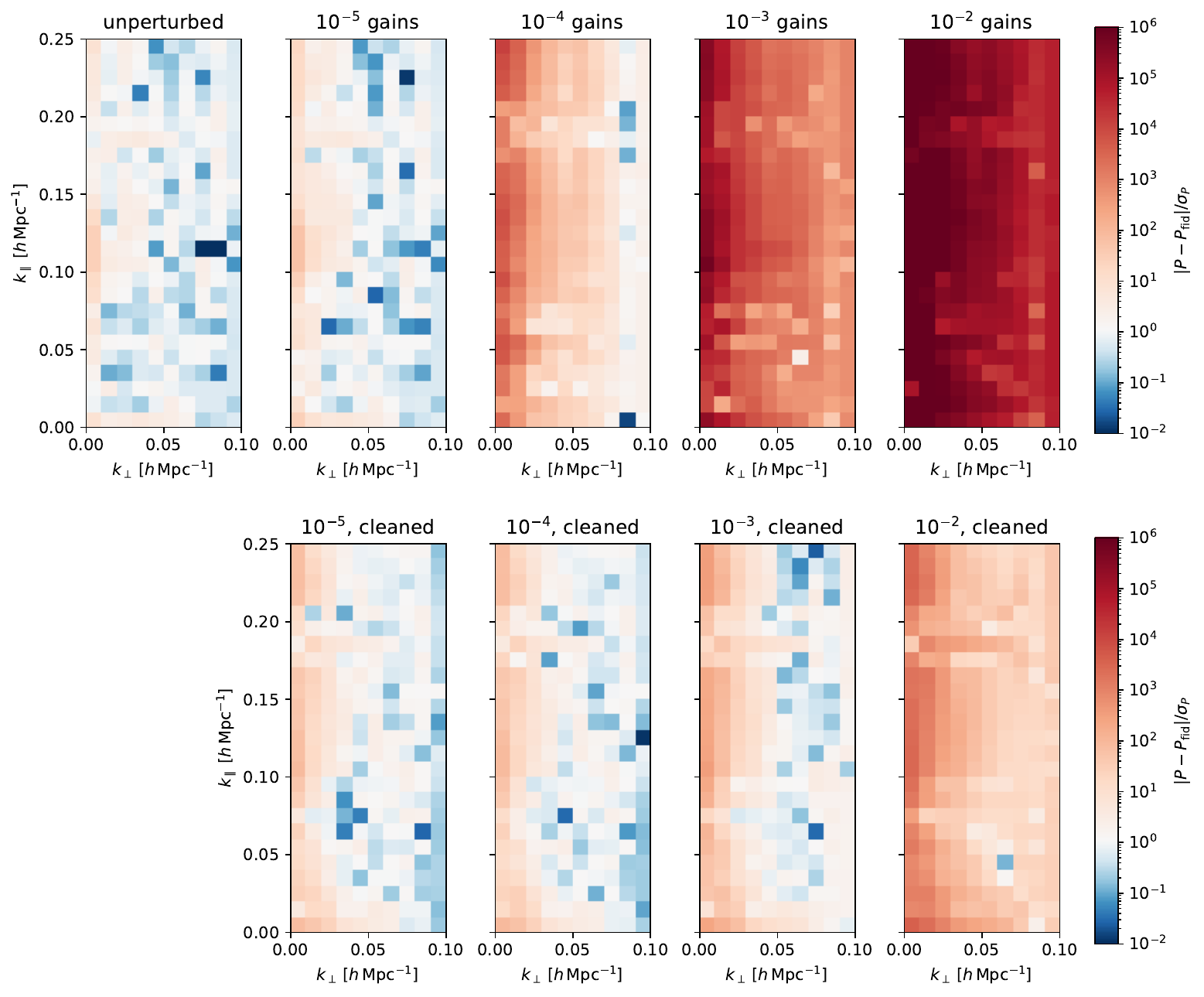}
\caption{Same as Fig.~\ref{fig:BP-pert} but implemented using the half-pathfinder setting with complex antenna-dependent errors. Due to sparse baseline sampling of the half-pathfinder telescope, foreground bias is seen at low $k_\perp$ without errors introduced. With antenna-dependent errors, HyFoReS is unable to suppress foreground bias to below the thermal noise level due to increased noise in the gain estimation, but HyFoReS nonetheless suppresses foreground bias by three orders of magnitude in the $10^{-3}$ and $10^{-2}$ cases.
\label{fig:AD-spec}}
\end{figure*}

\subsection{Complex antenna gains} \label{subsec:antenna_gains}

We then introduce complex antenna gain errors into the simulated data of the half-pathfinder telescope and compare the power spectra of KL-filtered visibilities, both with and without HyFoReS. The results are shown in Fig.~\ref{fig:AD-spec}. Unlike the case of the full pathfinder, we observe foreground bias of around one order of magnitude at low $k_\perp$ in the unperturbed power spectrum. This is due to the much sparser baseline sampling of the half-pathfinder telescope. 
As we increase the rms value of antenna gains from $10^{-5}$ to $10^{-2}$, we observe an increasing amount of foreground bias in the power spectra. After we implement HyFoReS, the foreground bias is reduced by three orders of magnitude for the gains of the $10^{-2}$ and $10^{-3}$ levels. However, unlike the case of bandpass errors, HyFoReS is unable to suppress foregrounds to below the thermal noise level for the $10^{-4}$ and $10^{-5}$ level antenna gains, leaving residual bias at low $k_\perp$. 

\begin{figure*}
\includegraphics[scale=0.64]{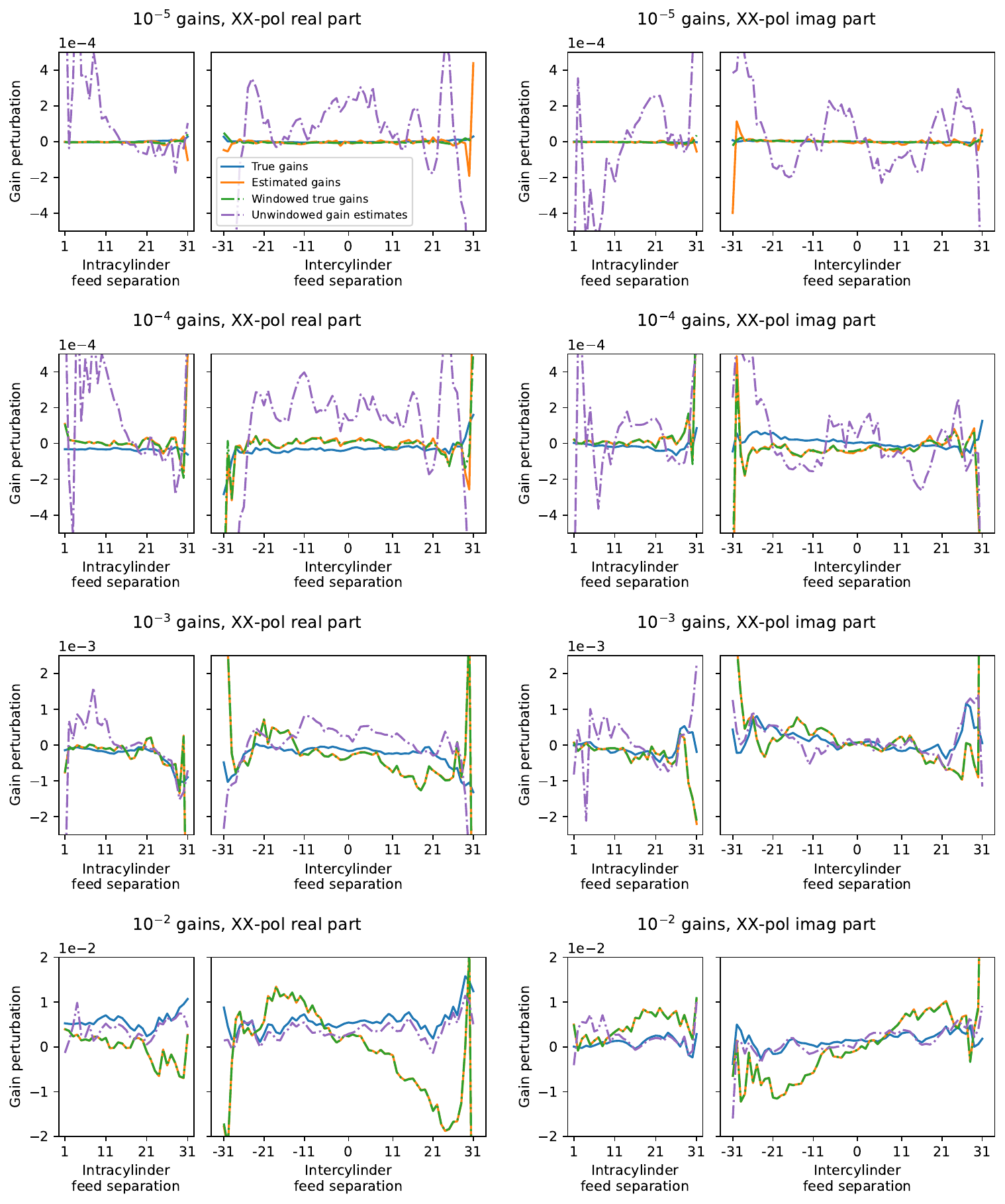}
\caption{Comparison between the true visibility errors from antenna-dependent gains (blue), estimated visibility errors (orange), true visibility errors passed through the window matrix (green), and unwindowed visibility error estimates (purple), plotted at 400 MHz as a function of north-south feed separations. Each panel is split into intracylinder and intercylinder baselines, corresponding to visibilities measured by antennas on the same cylinder and different cylinders, respectively. Large noise fluctuations are seen in the unwindowed gains in the $10^{-5}$ and $10^{-4}$ cases, indicating a relatively high noise level in the gain estimation. The noise becomes less dominant in the $10^{-3}$ and $10^{-2}$ cases as the gain errors increase. Only the XX polarization is shown, but the YY polarization is similar.\label{fig:comp_gains}}
\end{figure*}

To understand what limits the performance of the algorithm in this case, we recall that HyFoReS estimates the real and imaginary parts of the stacked visibility errors ($g_{\nu, bx}$ and $h_{\nu, bx}$ in Eq.~[\ref{eq: v_d_3}]) for each baseline $b$, polarization $x$ and frequency $\nu$. Figure~\ref{fig:comp_gains} shows the real and imaginary parts of the true visibility errors, the estimated errors ($\hat{g}_{\nu, bx}$ in Eqs.~[\ref{eq:g_hat_sim}] and $\hat{h}_{\nu, bx}$ in Eq.~[\ref{eq:h_hat_sim}]), the true errors passed through the window matrices (the first and second row of the right hand side of Eq.~[\ref{eq:g_hat_vec_real}]), and the unwindowed gain estimates ($\bm{\tilde{g}}$ and $\bm{\tilde{h}}$ from Eq.~[\ref{eq:g_tilde_real}]) at 400 MHz as a function of baseline feed separation. 

\begin{figure*}
\includegraphics[scale=0.5]{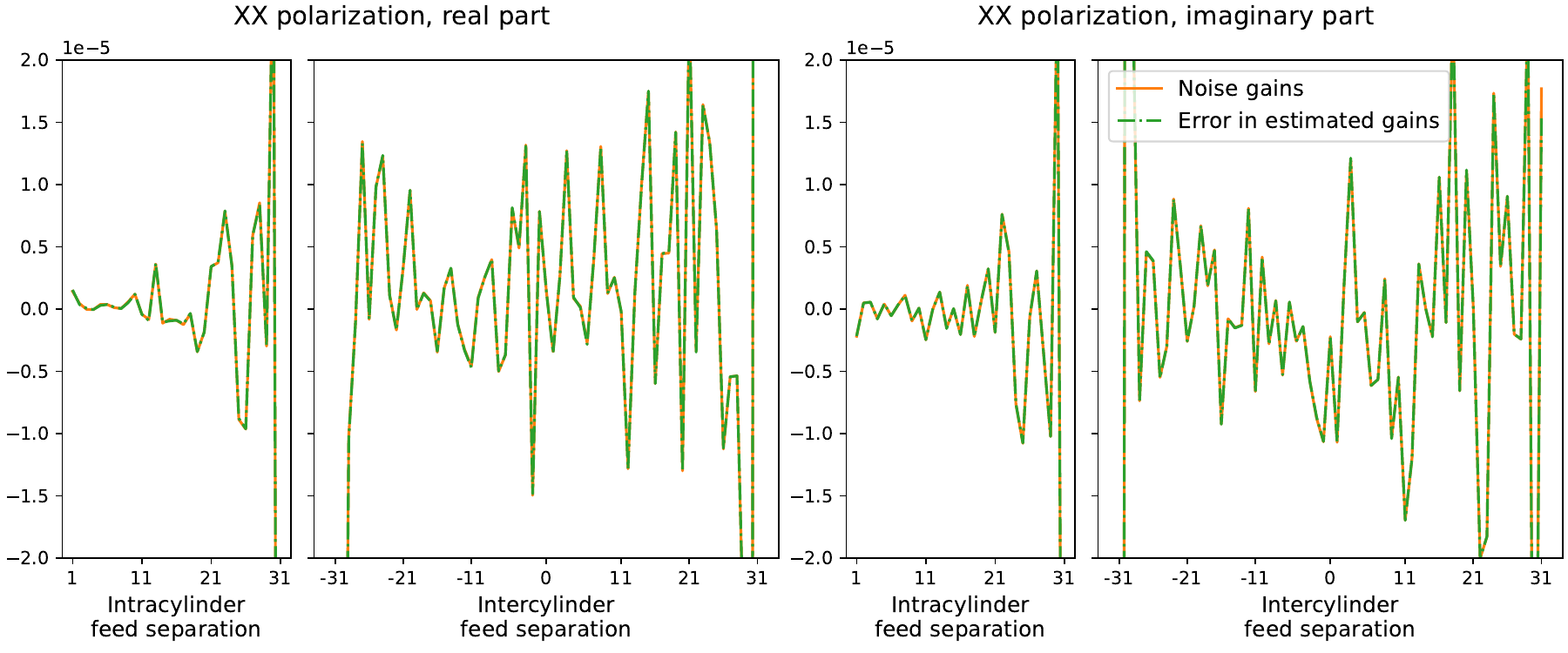}
\caption{Comparison between the error in the estimated visibility gains from the $10^{-4}$ case of Fig.~\ref{fig:comp_gains} (difference between the estimated gains and windowed true gains) and the predicted noise component of the estimated gains (noise gains). The matching curves indicate that the error in our gain estimates comes purely from spurious correlations between noise and the foregrounds. 
\label{fig:noise_gains}}
\end{figure*}

The most striking feature in Fig.~\ref{fig:comp_gains} is the large fluctuations of the unwindowed gains in the cases of $10^{-5}$ and $10^{-4}$ level antenna perturbations (purple curve, top 2 panels). 
This fluctuation comes from noise in the estimated gains (orange curve). To demonstrate this, note that the gain estimator in Eq.~(\ref{eq:window_form}) assumes that the leading term in the estimated signal is the foreground residuals $\sum_{i^\prime} \mbf{K}\mbf{\Gamma_{i^\prime}}\bm{v_F} g_{i^\prime}$. However, when the gains are small, such as at the $10^{-5}$ and $10^{-4}$ levels, the foreground residuals are weaker as well, so thermal noise in the estimated signal cannot be ignored. In this case, spurious correlations between the estimated foregrounds and noise (calling it $\bm{n_{d}}$) add to the right hand of Eq.~(\ref{eq:window_form}) as
\begin{align}
    \hat{y}_i & = \sum_{i^\prime} \frac{\bm{\hat{v}_F}^\dagger\mbf{E_i}^\dagger\mbf{K}\mbf{\Gamma_{i^\prime}}\bm{v_F}}{\bm{\hat{v}_F}^\dagger\mbf{D_i}\bm{\hat{v}_F}} g_{i^\prime} +  \frac{\bm{\hat{v}_F}^\dagger\mbf{E_i}^\dagger\mbf{K}\bm{n_{d}}}{\bm{\hat{v}_F}^\dagger\mbf{D_i}\bm{\hat{v}_F}} \label{eq:new_y_hat}\\
    & = \sum_{i^\prime} \mbf{W}_{i i^\prime}g_{i^\prime} + y^n_i, 
\end{align}
where we have defined
\begin{equation} \label{eq:noise_gains}
    y^n_i \equiv \frac{\bm{\hat{v}_F}^\dagger\mbf{E_i}^\dagger\mbf{K}\bm{n_{d}}}{\bm{\hat{v}_F}^\dagger\mbf{D_i}\bm{\hat{v}_F}}
\end{equation}
as the noise gains. To verify this term, we use the noise component of the simulated data and the estimated foregrounds to compute Eq.~(\ref{eq:noise_gains}). We then compare it with the error of the estimated gains, i.e., the difference between the estimated gains (orange curve in Fig.~\ref{fig:comp_gains}) and the windowed true gains (green curve in Fig.~\ref{fig:comp_gains}). 
The results are shown in Fig.~\ref{fig:noise_gains}. The agreement between the two quantities proves that the error in the estimated gains comes from spurious correlations between the estimated foregrounds and noise. 

To understand how noise in the estimated gains contributes to large fluctuations in the unwindowed gains, we recall that the unwindowed gains use the estimated gains and the window matrix to recover the frequency information in the original gains before KL filtering, including the low-pass modes suppressed by the KL filter. The low-pass noise components are magnified in this inversion process via the pseudo-inverted window, causing large noise fluctuations highly correlated across frequencies and baselines in the unwindowed gains. Fortunately, the effect of this low-pass noise on the final results is limited because when we use the unwindowed gains to subtract residual foregrounds, the KL filter is applied again (see Eq.~[\ref{eq:s_tilde_sim}]). This suppresses the low-pass components of the noise fluctuations. In summary, HyFoReS is able to subtract foreground residuals due to the antenna-dependent gains but does so at the cost of introducing multiplicative noise to the visibilities. This noise is then high-pass filtered, and the remaining high-pass noise is coupled with the foregrounds, introducing residual bias as seen in the lower panels of Fig.~\ref{fig:AD-spec}. 


Note that we did not observe significant noise in the estimated bandpass gains in the previous section. This is because bandpass gains are stationary over both the RA and baseline axes, over which the estimator $\mbf{E_i}$ of Eq.~(\ref{eq:noise_gains}) correlates the estimated foreground and noise. (In contrast, antenna gains are assumed to be stationary over the RA axis only.) The full pathfinder has almost 800 baselines, so the noise in the bandpass estimation is suppressed by a factor of $\sqrt{800}\approx 30$ (assuming Gaussian noise) compared to the noise in the antenna-dependent case. 


\begin{figure*}[htb!]
\includegraphics[width=2\columnwidth]{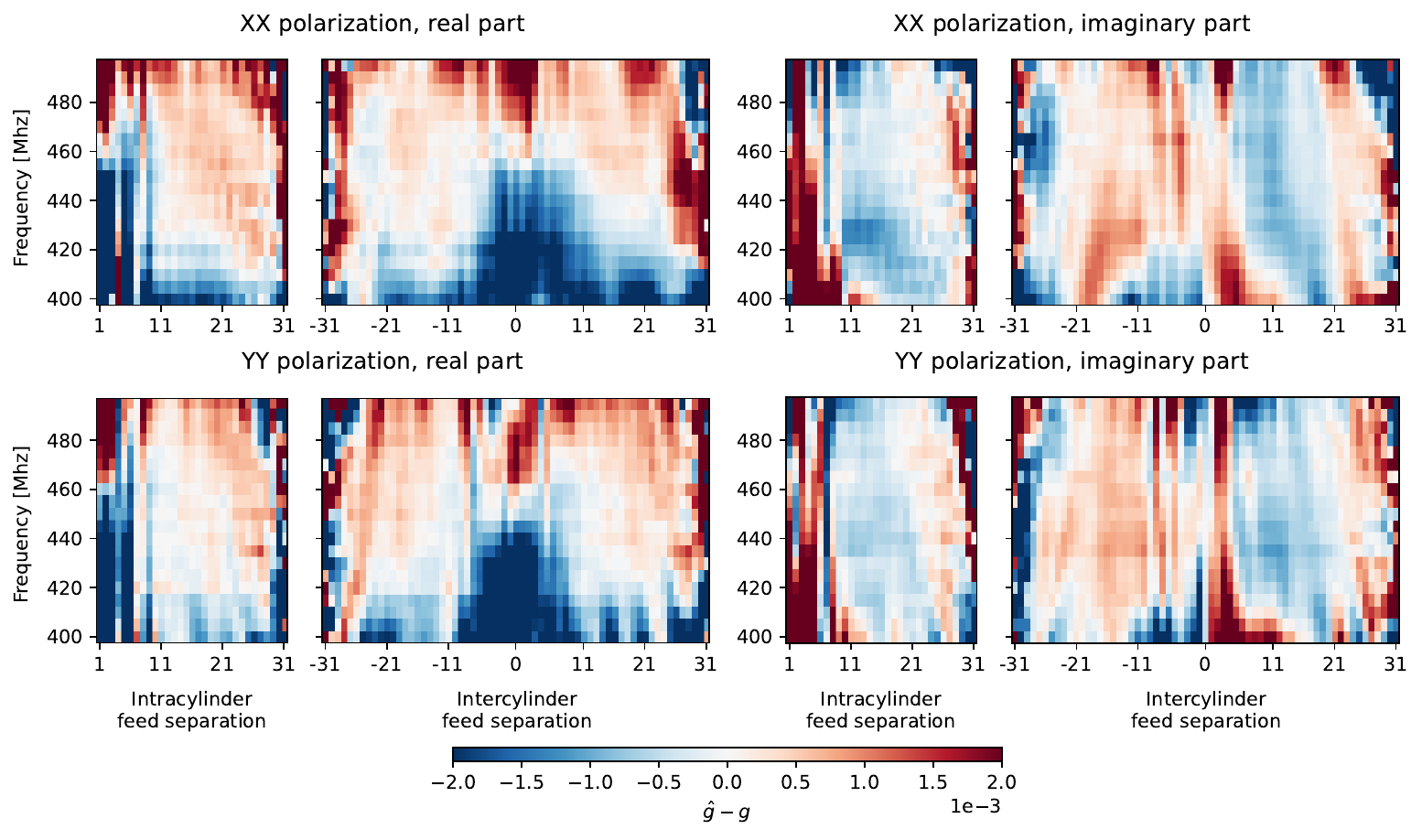}
\caption{The differences between the unwindowed visibility gain estimates and true visibility gains for the case of $10^{-2}$ level antenna-dependent errors as a function of frequency and north-south feed separation. The differences along the frequency axis correspond to the frequency modes removed by the KL foreground filter, and variations along the baseline direction arise from differences in the observed angular scales and noise levels of the stacked baselines.
\label{fig:ghat_g_1e-4}}
\end{figure*}

Lastly, for the case of $10^{-2}$-level antenna gains where the aforementioned noise is subdominant in the gains, we examine how different the unwindowed gains $\bm{\hat{g}}$ are from the true gains $\bm{g}$. We show their differences in Fig.~\ref{fig:ghat_g_1e-4} both on the frequency and baseline axes. Similar to the offset between the unwindowed and true gains in the bandpass case (Fig.~\ref{fig:BP-1e-3}), we find that $\bm{\hat{g}} - \bm{g}$ is smooth along the frequency axis at any baseline. This smoothness, again, matches our expectation of the frequency modes lost due to the KL filter. The differences across frequencies change with baselines because each baseline observes different angular scales on the sky and has different noise levels from redundant baseline stacking.

\section{discussion} \label{sec:dis}

\subsection{Comparing two types of gain perturbations} 

In this study, we used HyFoReS to mitigate gain-type errors in CHIME simulations. We tested two types of gain perturbations: bandpass errors on a full-pathfinder telescope, and complex antenna-dependent gains on a half-pathfinder telescope. We further assume that these gain errors are time independent. Comparing Figs.~\ref{fig:BP-pert} and \ref{fig:AD-spec}, we find that HyFoReS is more effective at mitigating bandpass errors than antenna-dependent errors, especially when the rms errors are below $10^{-3}$.

This is because antenna-dependent gains are much more complex than bandpass gains. The former, which contains both real and imaginary components, depends on frequency, baseline, and polarization, while the latter only varies with frequencies. The antenna-dependent perturbations thus have a much larger parameter space, meaning that there is much less redundant information in the data that we can use to estimate the gain parameters, leading to an increased amount of noise in the estimation. This is the main cause for the poorer performance in the case of antenna-dependent gains. Furthermore, the instrument setups are different for the two cases. The bandpass case is run with the full-pathfinder telescope, which measures more modes of the sky compared with the half-pathfinder telescope in the antenna-dependent case. This further contributes to the less ideal results in the antenna-dependent case.

Despite the simplicity of bandpass perturbations (i.e., having only frequency dependence), bandpass calibration is a critical and challenging part of real 21-cm experiments. We have shown that HyFoReS can reduce the foreground bias due to bandpass perturbations, thus relaxing the requirement on bandpass calibrations of real experiments. In reality, gain errors can vary with time, but observations of the sky are usually integrated over an extended time period. This can, in principle, mitigate time-dependent gain variations, although more careful analysis may be needed to understand the performance of HyFoReS on time-dependent systematics.


\subsection{Limitations of HyFoReS} \label{subsec:h_limit}

From Figs.~\ref{fig:BP-pert} and \ref{fig:AD-spec}, we find that HyFoReS can suppress gain-induced residual foreground bias by 3 orders of magnitude. However, the bias is not completely removed in some cases. We observe that the residual foreground bias is still above the noise level of the power spectrum for bandpass gains greater than $10^{-3}$ and in all cases of antenna-dependent errors. 

We identified two reasons for this. First, HyFoReS is currently designed to remove foreground residuals only due to systematic errors in their first order \citep{haochen_first_paper}. When gain errors are above the $10^{-3}$ level, second-order errors become nonnegligible since foreground residuals coupled with second-order perturbations will be brighter than the HI signal, i.e. $\mbf{G}^2\bm{v_{F}} \gtrsim \bm{v_{HI}}$. In the formalism of this study, second-order errors are introduced in Eq.~(\ref{eq:window_approx_form}) and similarly in Eqs.~(\ref{eq:W_RR}) through (\ref{eq:W_II}) when computing the window matrices. This is because the exact form of the window matrix, Eq.~(\ref{eq:window_def_form}) requires the knowledge of the true foregrounds, which we do not have. We instead use the data itself as an estimate for the foregrounds, which contains systematics, and thus introduce second-order errors. When we remove foreground residuals using Eq.~(\ref{eq:s_tilde_form}), only the foregrounds coupled with the first-order perturbations are subtracted. Lastly, ignoring cross-polarization visibilities in HyFoReS can also contribute to second-order errors (see Sec.~\ref{subsec:pol}).

Second, there is noise in the gain estimation (Eq.~[\ref{eq:new_y_hat}]). HyFoReS can then multiply this noise to visibilities during the foreground subtraction steps, introducing new foreground bias. We find that the noise in gain estimation comes from spurious correlations between foregrounds and noise in the data and can become significant if the parameter space for gain estimation is large.

Note that in real experiments such as CHIME, the antenna gains are already calibrated to a subpercent level \citep{chime_stacking}. Since the actual CHIME telescope has much better sensitivity than the simulated pathfinder telescopes, we expect HyFoReS to perform no worse than the results presented in this work, assuming the dominating source of systematics is time-independent gain errors. HyFoReS thus has the potential to reduce foreground bias due to gain errors in the real CHIME telescope data and to benefit the detection of the 21-cm auto-power spectrum.

\subsection{Limitations of CHIME simulations} \label{subsec:limit}

In this study, we use the CHIME software packages to simulate instruments similar to the CHIME pathfinder telescope. Although the simulated telescopes are much smaller than the full CHIME, our simulation nonetheless captures the main features of the real instrument. However, there are still notable issues in the current CHIME simulations. First, although we can simulate foreground emissions from point sources, the KL foreground filter in the pipeline does not include a point-source foreground model. This issue was discovered recently and is currently under investigation. As a result, we only simulate the Galactic synchrotron as the foreground. This makes our simulated foregrounds less realistic, but the Galactic synchrotron emission is the dominating foreground source for the 21-cm signal at large angular scales, and thus still serves as an effective tool to test HyFoReS. 

Second, due to the limitation of computational resources, we can only simulate a half-pathfinder telescope for antenna-dependent gains. The much sparser baseline sampling causes aliasing and foreground bias at low $k_\perp$ in the power spectrum. It is possible to improve the efficiency of the implementation of the HyFoReS algorithm so that we can afford to simulate a larger telescope in the future (see Sec.~\ref{subsec:improve}). Despite this limitation, the current simulation demonstrates the ability of HyFoReS to suppress gain-induced foreground bias.

Other systematics, such as the chromatic beam response, can also contribute to foreground leakage in the real telescope data \citep{chime_stacking}. HyFoReS was shown to remove beam-induced foreground residuals in \citep{wang2024} with a different implementation of the algorithm. Including more systematics and uncharacterized instrument responses in our simulations is an ongoing effort that will be further explored in future studies.

\subsection{Working with polarized data} \label{subsec:pol}

When HyFoReS was first developed in \citep{haochen_first_paper}, it was tested with complex antenna gain perturbations in a toy simulation. A key improvement of this study is to implement HyFoReS in a realistic simulation with polarized visibilities, which was never considered before. The pipeline in this study simulates data collected by dual-polarization antenna feeds, similar to the ones in the real CHIME experiment. When introducing antenna-dependent gain errors, we perturb both co-polarization and cross-polarization visibilities. However, when applying HyFoReS, we only estimate gain errors of the co-polarization and subtract foreground residuals only from the co-polarization visibilities (see Sec.~\ref{subsec:app}). This is because 
the foreground contribution to cross-polarization is much weaker than the foreground contribution to co-polarization. We therefore treat foreground contamination in cross-polarization as second-order perturbations. The results of Sec.~\ref{sec:result} show that this strategy is successful, although in future studies we can explore the possibility of including cross-polarization visibilities in foreground subtraction.




\subsection{Improvement in future studies} \label{subsec:improve}

We have demonstrated that HyFoReS can reduce foreground leakage due to gain errors in 21-cm data using realistic simulations of a CHIME-like telescope, but it is possible to improve the results in several ways. First, there is great potential to improve the implementation of the HyFoReS algorithm. The main reason for the high computational cost in the case of antenna-dependent gain errors is that the cross-correlation between the foreground and signal estimates (Eqs.~[\ref{eq:g_hat_sim}] and [\ref{eq:h_hat_sim}]) is performed in the KL space rather than the original visibility space. This requires us to forward model the effect of each gain error in the KL space, that is, computing $\mbf{K \Gamma_{\nu^\prime,b^\prime x^\prime}}$ in Eq.~(\ref{eq:E_R}) and $\mbf{K} \mbf{\Delta_{\nu,bx}}$ in Eq.~(\ref{eq:E_I}), for every baseline, frequency, and polarization. This operation alone takes $\sim\mathcal{O}(10^3)$ CPU hours in total, but can be completely skipped if we performed the cross-correlation in the original visibility space. Doing so requires an inverse KL transform that can map the KL-filtered visibilities ($\bm{\hat{v}_{HI}}$ for example) back to the original data space. We can add this inverse KL transform operation in the future to save computational cost and allow HyFoReS to run on larger simulations. In contrast, mitigating bandpass errors with HyFoReS is not as costly because of their low dimensionality. We are currently working on implementing HyFoReS on the real CHIME data to reduce the effect of bandpass errors and facilitate the detection of the 21-cm auto-power spectrum. 

To address the issue of second-order systematic errors, we could experiment applying HyFoReS iteratively. Namely, after having applied HyFoReS once, we cross-correlate the cleaned signal with the foreground estimate again to isolate and remove higher-order foreground residuals in the cleaned signal, although the increased level of complexity and non-linearity of the algorithm after multiple iterations may require careful analysis to control signal loss. To mitigate the spurious correlations between noise and foregrounds that we observed in Sec.~\ref{subsec:antenna_gains}, we could, in principle, use the form of the gain estimate and noise derived from Eq.~(\ref{eq:new_y_hat}) to construct a filter (for example, a Wiener filter) to statistically suppress noise-like modes in gain estimates. This may require detailed investigations in future studies.

There is also room for improvement in CHIME simulations. We are currently investigating the issues with the foreground models in the KL filter and are working on expanding the simulation to the full size of the CHIME telescope. More types of systematics, such as beam calibration errors, telescope focal line thermal expansions, and time-dependent gains, can be added to the simulations by ongoing work. We will consider how to use HyFoReS and/or other techniques to mitigate these systematics.

Lastly, HyFoReS is a general formalism that can be applied to any hydrogen intensity mapping experiment to address systematics-induced foreground leakage. In the future, we can explore implementing HyFoReS on other radio interferometers, such as those probing the 21-cm signal from the Epoch of Reionization or even other line intensity mapping experiments, to benefit the broader community.

\section{conclusion} \label{sec:con}

Systematics-induced foreground contamination is the biggest challenge for 21-cm intensity mapping experiments. In this work, we demonstrate using HyFoReS to remove foreground residuals induced by gain-type systematics in simulated CHIME data. We introduce bandpass and complex antenna-dependent gain perturbations to a full and half CHIME pathfinder-like telescope, respectively. The KL foreground filter in the default CHIME simulation pipeline cannot mitigate gain errors and leaves foreground residuals in the filtered data, even when the introduced gains are small with an rms value of $10^{-5}$. Increasing the rms values of the gains exacerbates residual foreground contamination, biasing the measured 21-cm power spectrum by up to 6 orders of magnitude when the perturbations are at the $10^{-2}$ level. 

We find that HyFoReS can suppress foreground residuals to mostly below the noise level of the power spectrum for $10^{-5}$ and $10^{-4}$ level bandpass errors. In comparison, the foreground subtraction results are less ideal for antenna-dependent errors, mainly due to noise in the gain estimation. Even when foregrounds are not completely subtracted, their bias is overall suppressed by three orders of magnitude in both bandpass and antenna-dependent gain cases. In real experiments such as CHIME, telescope gains are usually calibrated to a subpercent level. HyFoReS can then be applied to reduce the foreground bias caused by gain calibration errors to facilitate the detection of the cosmological 21-cm signal. 

Our study can be improved in a few ways, including making the HyFoReS algorithm implementation more efficient, scaling up the size of the simulated telescopes, building point-source foreground models into the KL filter, considering multiple sources of systematics, and improving foreground subtraction in the cross-polarization visibilities. Nonetheless, the simulation captures the main features of a realistic instrument and demonstrates the potential of using HyFoReS to mitigate systematics in 21-cm data. 

Our previous work \citep{wang2024} showed that HyFoReS suppressed beam-induced foreground residuals and improved 21-cm signal detection in the CHIME map. This study along with \citep{wang2024} establishes HyFoReS as a foreground subtraction algorithm that works in both simulations and real data, operates in either visibility or map space, and mitigates different types of systematics. We hope that HyFoReS adds to the arsenal of tools for mitigating foregrounds and telescope systematics, thus moving forward the field of line intensity mapping experiments.

\begin{acknowledgements}
We thank Richard Shaw and Seth R. Siegel for their work on building the CHIME simulation pipeline. We thank  Juan Mena-Parra for useful discussions on this study. We thank the Dominion Radio Astrophysical Observatory, operated by the National Research Council Canada, for gracious hospitality and expertise. The DRAO is situated on the traditional, ancestral, and unceded territory of the Syilx Okanagan people. We are fortunate to live and work on these lands.

CHIME is funded by grants from the Canada Foundation for Innovation (CFI) 2012 Leading Edge Fund (Project 31170), the CFI 2015 Innovation Fund (Project 33213), and by contributions from the provinces of British Columbia, Québec, and Ontario. Long-term data storage and computational support for analysis is provided by WestGrid, SciNet and the Digital Research Alliance of Canada, and we thank their staff for flexibility and technical expertise that has been essential to this work.

Additional support was provided by the University of British Columbia, McGill University, and the University of Toronto. CHIME also benefits from NSERC Discovery Grants to several researchers, funding from the Canadian Institute for Advanced Research (CIFAR), from Canada Research Chairs, from the FRQNT Centre de Recherche en Astrophysique du Québec (CRAQ) and from the Dunlap Institute for Astronomy and Astrophysics at the University of Toronto, which is funded through an endowment established by the David Dunlap family. This material is partly based on work supported by the NSF through grants (2008031) (2006911) (2006548) and (2018490) and by the Perimeter Institute for Theoretical Physics, which in turn is supported by the Government of Canada through Industry Canada and by the Province of Ontario through the Ministry of Research and Innovation.

K.W.M. holds the Adam J. Burgasser Chair in Astrophysics and is supported by NSF grants (2008031, 2018490).
\end{acknowledgements}



\bibliography{journals,lit}
\end{document}